\documentclass[a4paper]{article}
\usepackage{graphicx}
\usepackage{subeqn}
\newcommand{\U}{{\mathcal U}}
\newcommand{\Z}{{\mathcal Z}}

\begin{document}

\begin{center}
{\Large FIRST ORDER PHASE TRANSITION IN A MODEL FOR GENERALISED STATISTICS}
\bigskip \bigskip

\small


DRAGO\c S-VICTOR ANGHEL

\bigskip

\footnotesize

{\em National Institute for Nuclear Physics and Engineering, P.O.Box MG-6,\\ RO-077125 Bucharest-Magurele, Romania, E-mail: dragos@theory.nipne.ro}

\bigskip

\small

(Received \today ) 

\end{center}

\bigskip

\footnotesize

{\em Abstract.\/} 
A first order phase transition is found in a model which was introduced 
originally by Murthy and Shankar [Phys. Rev. B {\bf 60}, 6517 (1999)] to 
describe systems of generalised exclusion statistics. 
I characterise the phase transition in the 
canonical and grandcanonical ensebles 
for the case when the statistical exclusion parameter is 1, which corresponds 
to the Fermi exclusion statistics. We observe that in the grandcanonical 
ensemble the phase transition has no latent heat, but it has a finite 
jump in the particle number. In canonical conditions--when 
the particle number is held fix--the internal energy is discontinuous at 
the transition. 

\bigskip


\normalsize

\section{INTRODUCTION \label{Intro}}

The exclusion statistics that it is intermediate between the Bose and 
the Fermi statistics, called \textit{fractional exclusion statistics} (FES), 
was introduced by Haldane in Ref. \cite{PhysRevLett.67.937.1991.Haldane} to 
describe, among other systems, quasiparticle excitations in fractional 
quantum Hall effect and spinon 
excitations in spin-$\frac{1}{2}$ quantum ferromagnets. For this he 
assumed that the many-particle system has a finite-dimensional Hilbert 
space whos dimension varies linearly with the particle number in the system. 

The thermodynamic properties of systems with FES in the thermodynamic limit 
were clarified mainly by Isakov \cite{PhysRevLett.73.2150.Isakov} and 
Wu \cite{PhysRevLett.73.922.1994.Wu}. 

One of the interacting models that leads to FES was introduced first by 
Murthy and Shankar in Ref. \cite{PhysRevLett.73.3331.1994.Murthy} and 
then used with slight variations in different 
contexts by various authors (see for example \cite{JPhysB33.3895.2000.Bhaduri,PhysRevLett.86.2930.2001.Hansson,IntJModPhysA12.1895.1997.Isakov,JPA35.7255.2002.Anghel}). 
To explain the model, let us assume for the beginning that we have an ideal 
gas of bosons, with the single particle levels denoted by $\epsilon_i$, 
$i=0,1,\ldots$ and $\epsilon_i\le\epsilon_{i+1}$ for any $i$. 
I also assume, for the sake of clarity, that the energy levels are 
equidistant and I introduce the \textit{density of states}, 
$\sigma\equiv(\epsilon_{i+1}-\epsilon_i)^{-1}$. 
The interaction between particles changes the energy levels in such a way 
that the energy level $i$\
changes its energy to 
\begin{equation}
\tilde\epsilon_i\equiv\tilde\epsilon(\epsilon_i) = 
\epsilon_i+\sigma^{-1}gN_{i-}+ \sigma^{-1}h_0\max(n_i-1,0) \,, 
\label{interaction0} 
\end{equation}
where $n_i$ is the numer of particles on the level $i$, while 
$N_{i-}\equiv\sum_{j<i} n_j$ is the number of particles below the level 
$i$. By $\max(a,b)$ I denote the maximum of $a$ and $b$. The constants 
$g$ and $h_0$ are adimensional interaction parameters. 
In order to ensure the stability of the 
system, I impose from the beginning $h_0\ge 0$ and $g\ge 0$. 
Obviously, $h_0=g=0$ corresponds to the ideal Bose gas. 

Let us now analyse the (quasi)continous limit. In this limit 
we can relax the condition of having an equidistant single particle spectrum 
and describe the system only by the \textit{average} density of states 
$\sigma$. Then for any $\epsilon>0$, if there are no macroscopically 
populated energy levels--i.e. in the absence of condensation--, 
$n(\epsilon)$ is much smaller than $N_{\epsilon -}$ and we can write
\begin{equation}
\tilde\epsilon(\epsilon) = \epsilon+ g\int_0^\epsilon n(\epsilon)
d\epsilon \,. \label{interaction0qc} 
\end{equation}
Even for $\epsilon=0$ we can neglect the microscopic offset 
$\sigma^{-1}h_0n(\epsilon=0)$ of the groundstate--which has no contribtion 
to the thermodynamics of the system--and write $\tilde\epsilon(0)=0$. 

Now we can prove the generalised exclusion statistics character 
of the system. For this, I will take the arbitrary interval 
$[\tilde\epsilon_{\rm l},\tilde\epsilon_{\rm u}]$, which contains 
a large enough number of particles, 
$n_{[\tilde\epsilon_{\rm l},\tilde\epsilon_{\rm u}]}$, and 
calculate the number of states it contains. The energies 
$\tilde\epsilon_{\rm l}$ and $\tilde\epsilon_{\rm u}$ are related to 
the energy levels in the noninteracting system, $\epsilon_{\rm l}$ and 
$\epsilon_{\rm u}$, by the relations
\begin{eqnarray}
\tilde\epsilon_{\rm l} &=& \epsilon_{\rm l} +\sigma^{-1} 
[gN_{\tilde\epsilon_{{\rm l}-}}+h_0 \max(n_{\tilde\epsilon_{\rm l}},0)] \,, 
\label{tilde_epsl_epsl} \\
\tilde\epsilon_{\rm u} &=& \epsilon_{\rm u} +\sigma^{-1} 
[g(N_{\tilde\epsilon_{{\rm l}-}} 
+n_{[\tilde\epsilon_{\rm l},\tilde\epsilon_{\rm u}]}) 
+h_0 \max(n_{\tilde\epsilon_{\rm u}},0)] \,. \label{tilde_epsu_epsu} 
\end{eqnarray}
For enough many energy states in the interval 
($n_{[\tilde\epsilon_{\rm l},\tilde\epsilon_{\rm u}]}\gg n_{\tilde\epsilon_{\rm l}},\ n_{\tilde\epsilon_{\rm u}}$) I can write the number of states 
between $\tilde\epsilon_{\rm l}$ and $\tilde\epsilon_{\rm u}$ as 
\begin{equation} 
d_{[\tilde\epsilon_{\rm l},\tilde\epsilon_{\rm u}]} = 
\sigma(\epsilon_{\rm u}-\epsilon_{\rm l}) = 
\sigma(\tilde\epsilon_{\rm u}-\tilde\epsilon_{\rm l})
-n_{[\tilde\epsilon_{\rm l},\tilde\epsilon_{\rm u}]}g \,.
\label{in_num_states}
\end{equation}
Relation (\ref{in_num_states}) is linear in the number of particles in the 
interval and gives an \textit{effective exclusion statistics parameter} 
$\alpha=g$ \cite{JPhysB33.3895.2000.Bhaduri,PhysRevLett.86.2930.2001.Hansson,IntJModPhysA12.1895.1997.Isakov,JPA35.7255.2002.Anghel}. 
The case $\alpha=0$ corresponds to Bosons, whereas $\alpha=1$ corresponds 
to Fermions. Therefore the interacting gas is described as an 
``ideal'' gas of particles of generalised statistics, with the 
\textit{same}, constant density of states $\sigma$. 

The thermodynamics of such a system is obtained by splitting the energy 
axis into intervals which are small enough, so that the energy variations 
caused by different rearrangements of particles in any of the intervals are 
insignificant, but which are also sufficiently large, so that each interval 
contains a 
large number of particles and energy levels. If we number these energy 
intervals by capital letters 
(e.g. $I=0,1,\ldots$) and the interval $I$ contains $\delta n_I$ 
particles and $d_I$ energy levels, then the number of microstates obtained 
by rearranging the particles in this interval is 
\begin{equation}
w_i = \frac{(\delta n_I+d_I)!}{\delta n_I!d_I!} \label{nmb_microstates}
\end{equation}
Summing up the number of microstates, multiplied by their statistical weights, 
one obtains the partition function, from which, by taking the logarithm 
and using the Stirling formula, one calculates the thermodynamic potential 
and entropy (see for example \cite{PhysRevLett.73.2150.Isakov,PhysRevLett.73.922.1994.Wu,PhysRevB.60.6517.1999.Murthy}). The thermodynamic potentials 
have been put into an universal form for any $\alpha$ by writing them in 
terms of polylogarithm functions \cite{JPA35.7255.2002.Anghel}, as 
previously done independently for Bosons ($\alpha=0$) and Fermions 
($\alpha=1$) by Lee \cite{PhysRevE.55.1518.1997.Lee} and Viefers, 
Ravndal, and Haugset\cite{AmJPhys63.369.Viefers}. For example the 
grandcanonical potential, $\Omega$, and the entropy, $S$, can be written 
in the form 
\begin{subequations}\label{fracUniv}
\begin{eqnarray}
\Omega &=& -U = \frac{1-\alpha}{2}\frac{N^2}{\sigma} + 
(k_{\rm B} T)^2 \sigma Li_2(-y_0) , \label{echiv} \\
S &=& -k_{\rm B}^2 T \sigma [2Li_2(-y_0) + \log{(1+y_0)}\log{y_0}] \,,
\label{entropia} 
\end{eqnarray}
\end{subequations}
if we define $y_0$ by the equation 
$(1+y_0)^{1-\alpha}/y_0 = e^{-\mu/k_{\rm B}T}$. In Eq. (\ref{echiv}) 
$N$ is the total particle number in the system and $\mu$ is the 
chemical potential. 

Equation (\ref{entropia}) shows that 
the entropy of any system of constant DOS is independent 
of the exlcusion statistics parameter $\alpha$ \cite{JPA35.7255.2002.Anghel}. 
From here follows that in canonical conditions 
the systems of any exclusion statistics, but of the same, constant density 
of states, are identical from 
the thermodynamic point of view. This property has been given the name of 
\textit{thermodynamic equivalence} of systems of constant DOS 
\cite{PhysRevE.55.1518.1997.Lee}. 

\subsection{THERMODYNAMIC EQUIVALENCE \label{ThEq}}

Apparently $h_0$ plays no role in the model since does not enter any of the 
macroscopic equations (\ref{in_num_states}), (\ref{nmb_microstates}), or 
(\ref{fracUniv}). For this reason it has been in general 
neglected \cite{IntJModPhysA12.1895.1997.Isakov}, if not entirely 
discarded \cite{PhysRevLett.73.3331.1994.Murthy,JPhysB33.3895.2000.Bhaduri,PhysRevLett.86.2930.2001.Hansson,PhysRevB.60.6517.1999.Murthy,PhysRevLett.74.3912.1995.Sen}. 
After all, a thermodynamic Bose system is usually thought of as an 
infinite Bose gas, with $\sigma\to\infty$, $N\to\infty$, and finite 
density $\rho\equiv N/\sigma$. Such a gas does not condense at any 
finite temperature, so there is no reason to consider $h_0$. 

Moreover, the model can be built from a fermionic perspective. In 
\cite{JPA35.7255.2002.Anghel} I showed that the microscopic reason 
for the thermodynamic equivalence between a Bose and a Fermi system 
of the same, constant density of states is the possibility to make a 
one-to-one mapping between the microscopic configurations of bosons and 
the microscopic configurations of fermions, with the same excitation 
energy \cite{submitted.EST}. By this mapping, the bosons that lay on the same 
energy level are transformed into a goup of ``close packed'' fermions, 
i.e. into a group of fermions which occupy all the states into a 
single-particle energy interval (more details are given in Refs. 
\cite{JPA35.7255.2002.Anghel,submitted.EST}). In such 
a case, the single particle energies of the  interacting Fermi gas 
corresponding to Eq. (\ref{interaction0}) should be written as 
\begin{equation}
\tilde\epsilon_i\equiv\tilde\epsilon(\epsilon_i) = 
\epsilon_i+\sigma^{-1}g^{(F)} N_{i-}+ \sigma^{-1}h^{(F)}_0\max(n^{(h)}_I-1,0) 
\,, \label{interaction1} 
\end{equation}
with $g^{(F)}=1-g$ and $h^{(F)}_0=h_0-0.5$. In this description, 
$n^{(F)}_I$ is the number of fermions in the ``close-packed'' group that 
contains the particle on the level $i$. Since at temperature $T$ in any 
small energy interval $\delta\epsilon$, centered at $\tilde\epsilon$, there 
are on average 
$\sigma\delta\epsilon/\{1+\exp[\beta(\mu-\tilde\epsilon)]\}\ (\ne 0)$ holes, 
we cannot expect that a macroscopical interval along the energy axis will be 
``close-packed,'' so from this perspective also--maybe even more 
clearly--the parameter $h_0$ does not appear to have any macroscopic 
significance. 

\subsection{THE ROLE OF $h_0$ \label{roleh0}}

Infinite systems of constant density of states are indeed not condensed 
at any finite temperature, as stated before. Nevertheless, in finite systems, 
a significant fraction of particles may accumulate on the ground state 
at low enough temperature in Bose systems, or it can arrange in a 
close-packed (or degenerate) configuration on the lowest energy levels 
in Fermi systems \cite{JPA36.L577.2003.Anghel}. Therefore, at low enough 
temperature we should take into account separately the population of the 
ground state in Bose systems or the corresponding degenerate fermionic 
subsystem and I shall refer to them as the Bose or the Fermi condensates 
\cite{JPA36.L577.2003.Anghel}. 

By doing so, in Ref. \cite{JPA35.7255.2002.Anghel} I showed that if 
$h_0<0.5$ the system undergoes a first order phase transition for any 
$\alpha$. In this paper I will describe in detail the phase transition 
for $\alpha=1$. In Section \ref{PhTrSec} I calculate 
the partition functions and I show how the phase transition 
occures in grandcanonical (Section \ref{PhTrGC}) and canonical systems 
(Section \ref{PhTrC}). 
In Section \ref{PhTrDisc} I calculate the discontinuities that 
appear in the thermodynamical functions at the phase transitions. I 
shall show that in the grandcanonical ensemble the internal energy 
is continuous at the transition, while the particle number has a jump. 
However, in the canonical ensemble, since the particle number is conserved, 
the internal energy is discontinuous at the phase transition. Last section 
is reserved for conclusions. 

\section{THE PHASE TRANSITION \label{PhTrSec}}

\subsection{THE GRANDCANONICAL ENSEMBLE \label{PhTrGC}}

From now on I shall discuss the case $\alpha=g=1$. Although the 
analysis is equally easy from Bose and Fermi perspectives, for 
easier reference to the calculations in \cite{JPA35.7255.2002.Anghel} 
I shall adopt here the bosonic picture. 

To emphasise the condensate, I write the grandcanonical partition function as 
a sum of terms corresponding to different values of $n_0$ (the 
population of the condensate) \cite{JPA35.7255.2002.Anghel}, 
\begin{equation}
\Z = \sum_{n_o}\Z_{n_0}\equiv\sum_{n_o}e^{-\beta\frac{h_0 n_0^2}{\sigma}+ 
\beta\mu n_0}\cdot\prod_{i=1}^\infty e^{-\beta(\tilde\epsilon_i-\mu)n_i} 
\equiv \sum_{n_o}e^{-\beta\frac{h_0 n_0^2}{\sigma}+\beta\mu n_0}\cdot
\Z_{\rm ex}(n_0,\beta)
\,. \label{label_partition1}
\end{equation}
The ``partial'' partition function, $\Z_{n_0}$, is proportional 
to the probability that a configuration with $n_0$ particles in the 
condensate occures. As it happens usually at a first order phase transition, 
the partition function in some parameter space develops two maxima. 
In each of the phases one maximum is dominant and the equilibrium 
satate of the system corresponds to it. As we change from one phase 
to the other by varying some of the thermodynamic parameters of the system 
(e.g. the temperature or the chemical potential), 
the difference between the two maxima decreases and they become equal 
at the phase transition. Moving further into the second phase, the maximum 
corresponding to it becomes dominant. For our system, the parameter with 
which we describe the phase transition is the population of the 
condensate, $n_0$. 

Therefore, instead of calculating the whole partition function, 
I calculate $\Z_{n_0}$ and analyse its maxima. It is easier 
to work with  $\ln\Z_{n_0}$, which is 
\begin{equation}
\ln\Z_{n_0} = -\beta\frac{h_0 n_0^2}{\sigma}+\beta\mu n_0 + 
\ln\Z_{\rm ex}(n_0,\beta). \label{ln_Zn0}
\end{equation}
I introduce the dimensionless quantities $a\equiv\beta\mu$, 
$\xi\equiv\beta N/\sigma$, and $\xi_0\equiv\beta n_0/\sigma$. In these 
notations Eq. (\ref{ln_Zn0}) becomes 
\begin{equation}
\beta\sigma^{-1}\ln\Z_{n_0} = -h_0\xi_0^2 + a\xi_0 + 
\beta\sigma^{-1}\ln\Z_{\rm ex}(n_0,\beta) \,. \label{ln_Zn0_xi}
\end{equation}
According to the calculation procedure outlined in Section \ref{Intro}, 
$\Z_{\rm ex}(n_0,\beta)$ is the partition function of a Fermi gas with 
the same chemical potential and with a single particle spectrum that 
starts at $\tilde\epsilon_1$: 
\begin{equation}
\ln\Z_{\rm ex}(n_0,\beta) = \int_{\tilde\epsilon_1}^\infty\ln\left(
1+e^{a-\beta\epsilon}\right)\sigma d\epsilon . \label{ln_Zex}
\end{equation}
In Ref. \cite{JPA35.7255.2002.Anghel} I showed that condensation may occure 
only on the first energy level, so I can write 
$\tilde\epsilon_1=\epsilon_1+n_0\sigma^{-1}\equiv\sigma^{-1}(1+n_0)$, 
which leads to 
\begin{equation}
\beta\sigma^{-1}\ln\Z_{\rm ex}(n_0,\beta) = 
\int_{\sigma^{-1}(1+n_0)}^\infty\ln\left(1+e^{a-\beta\epsilon}\right) 
\beta d\epsilon = \int_{\beta\sigma^{-1}+\xi_0}^\infty
\ln\left(1+e^{a-x}\right)dx \,.
 \label{ln_Zex1}
\end{equation}
I plug this into (\ref{ln_Zn0_xi}) and I get 
\begin{eqnarray}
\beta\sigma^{-1}\ln\Z_{n_0} &=& -h_0\xi_{0}^2 + a\xi_{0} + 
\int_{0}^\infty\ln\left(1+e^{a-\xi_0-\beta\sigma^{-1}-x}\right)dx \nonumber \\
&\equiv& -h_0\xi_{0}^2 + a\xi_{0} - Li_2(-e^{a-\beta\sigma^{-1}-\xi_0}) \,, 
\label{ln_Zn0_xi0}
\end{eqnarray}
where I used the notation 
\begin{equation}
Li_n(z)\equiv \frac{1}{\Gamma(n)}\int_0^\infty \frac{x^{n-1}\,dx} 
{e^{x}z-1} = \frac{z}{1^n}+\frac{z^2}{2^n}+\frac{z^3}{3^n}+\ldots
\end{equation}
for the polylogarithmic function of order $n$ 
\cite{PhysRevE.56.3909.1997.Lee,Lewin:book}. 

Now we have to find and compare the maxima of $\ln\Z_{n_0}$. 
If $\ln\Z_{n_0}$ forms two maxima, then our system has two phases. In 
each of the phases one maximum is dominant and $n_0$, together with 
all the measurable thermodynamic quantities, take the values corresponding 
to this maximum \cite{PhysRevE53.6558.1996.Lee,PhysRevE62.4558.2000.Lee,PhysRevLett84.4794.2000.Biskup,PhysRevE64.046114.2001.Chomaz}. 
Since $\ln\Z_{n_0}$ is a smooth function, we find its extrema at $n_0>0$ by 
equating to zero its derivatives with respect to $n_0$ or, more conveniently, 
with respect to $\xi_0$: 
\begin{eqnarray}
0 &=& \frac{\partial(\beta\sigma^{-1}\ln\Z_{n_0})}{\partial \xi_0} = 
-2h_0\xi_0+a-\ln\left(1+e^{a-\beta\sigma^{-1}-\xi_0}\right) \label{Eq_n0max1}
\end{eqnarray}
We transform it a bit, to obtain 
\begin{equation}
\beta\sigma^{-1}+\xi_0(1-2h_0)=\ln\left[1+e^{-(a-\xi_0)+\beta\sigma^{-1}}
\right] \label{Eq_n0max2}
\end{equation}
For an ideal Fermi gas, $h_0=1/2$ and we reobtain the equation for the 
Fermi condensate in ideal gases of constant $\sigma$ 
\cite{JPA36.L577.2003.Anghel}: 
\begin{equation}
\beta\sigma^{-1}=\ln\left[1+e^{-(a-\xi_0)+\beta\sigma^{-1}}
\right] \label{HA!}
\end{equation}
In general, if $h_0>1/2$, the derivative 
$\partial(\beta\sigma^{-1}\ln\Z_{n_0})/\partial \xi_0$ is always negative and 
$\Z_{n_0}$ has only one maximum at $n_0=0$. We have therefore only one 
(uncondensed) phase for any $\alpha$ \cite{JPA35.7255.2002.Anghel}. 

\begin{figure}
\begin{center}
\resizebox{100mm}{!}{\includegraphics{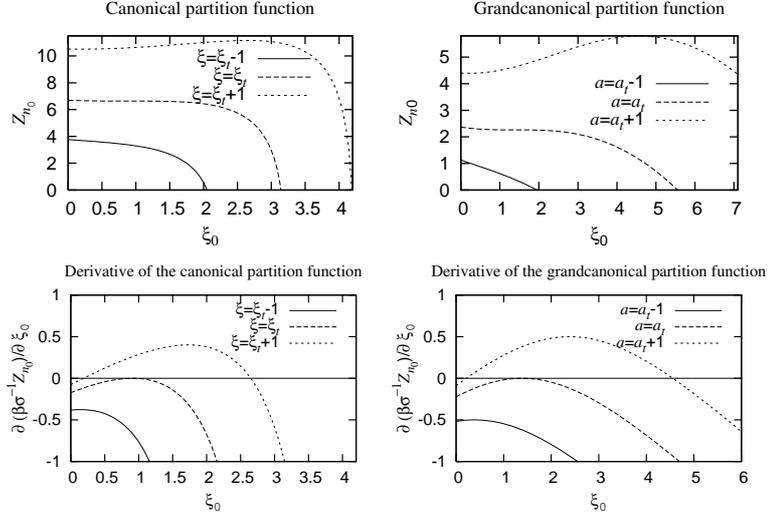}}
\end{center}
\caption{The partition functions of the canonical and grandcanonical 
ensembles and their derivatives}\label{Zn0_Zn0der}
\end{figure}

If $h_0<1/2$ we have a condensation \cite{JPA35.7255.2002.Anghel}. 
In this case, since the condensate contains 
a macroscopic number of particles, we can neglect $\beta/\sigma$ as 
compared to $a$ and $\xi_0$ and write Eq. (\ref{Eq_n0max2}) in the 
shorter form: 
\begin{equation}
\xi_0(1-2h_0)=\ln\left[1+e^{\xi_0-a}\right] \label{Eq_n0max3}
\end{equation}
Equation (\ref{Eq_n0max3}) is easy to interpret. For small enough $a$, 
the functions $\xi_0(1-2h_0)$ and $\ln[1+\exp(\xi_0-a)]$ do not 
cross each-other, so we have no solution for (\ref{Eq_n0max3}) and 
$\Z_{n_0}$ has only one maximum at $n_0=0$--no condensate. This situation 
is plotted with continuous lines in the right upper and lower plots 
of Fig. \ref{Zn0_Zn0der}. 

As $a$ increases, $\xi_0(1-2h_0)$ 
becomes tangent to $\ln[1+\exp(\xi_0-a)]$ when 
\[
\frac{d}{d\xi_0}\ln[1+e^{\xi_0-a}] = \frac{1}{1+e^{a-\xi_0}} = 1-2h_0 \,,
\]
which happends at $\xi_{0,t}=a-\ln[2h_0/(1-2h_0)]$. Plugging $\xi_{0,t}$ 
into Eq. (\ref{Eq_n0max3}) I get 
\begin{equation}
a_t = -\frac{2h_0\ln(2h_0)+(1-2h_0)\ln(1-2h_0)}{1-2h_0} \,. \label{at}
\end{equation}
This situation is drawn in dashed lines in the right-hand plots of 
Fig. \ref{Zn0_Zn0der}. 

For $a>a_t$, Eq. (\ref{Eq_n0max3}) has two solutions, $\xi_{0,1}$ 
and $\xi_{0,2}$, with $\xi_{0,1}<\xi_{0,2}$. This means that 
for $\xi_0<\xi_{0,1}$ the partition function $\Z_{n_0}$ decreases 
with increasing $n_0$, for $\xi_{0,1}<\xi_0<\xi_{0,2}$, $\Z_{n_0}$ increases 
with $n_0$, and finally for $\xi_0>\xi_{0,2}$, $\Z_{n_0}$ decreases again. 
The two competing maxima of $\Z_{n_0}$, one at $n_0=0$ and the other one at 
$n_0=k_BT\sigma\xi_{0,2}$, indicate a potential phase transition. 
Indeed, as $a$ increases past $a_t$, the relative height of the 
maximum located at $\xi_0=\xi_{0,2}$ increases with respect to the 
maximum at $\xi_0=0$, and at a certain critical value, $a_{cr}$, we obtain 
$\Z_{n_0=0}=\Z_{n_0=k_BT\sigma\xi_{0,2}}$. Above $a_{cr}$ the second 
maximum is higher (see the dotted lines in the plots of Fig. 
\ref{Zn0_Zn0der}), $\Z_{n_0=0}=\Z_{n_0=k_BT\sigma\xi_{0,2}}$, and we 
say that a phase-transition occured. As we shall see in Section \ref{DiscGC}, 
the (average) internal energy of the gas is conserved at the transition, but 
the particle number has a jump. 

This is a first order phase transition. 

\subsection{THE CANONICAL ENSEMBLE \label{PhTrC}}

Let us now analyse the system in canonical conditions. 
Neglecting again the microscopic energy step $\sigma^{-1}$ in the 
system with interaction, the total particle number is 
\begin{eqnarray}
N &\equiv& n_0+N_{\rm ex} = n_0+\int_{\tilde\epsilon_1}^\infty 
\frac{\sigma d\epsilon}{e^{\beta\epsilon-a}-1} 
\approx n_0+ k_B T\sigma\ln\left(1+e^{a-\xi_0}\right) \label{Nex}
\end{eqnarray}
If I denote $\xi_{\rm ex}\equiv \beta\sigma^{-1}N_{\rm ex}$, then 
\begin{equation}
\xi_{\rm ex} = \ln\left(1+e^{a-\xi_0}
\right) \label{xiex}
\end{equation}
and Eq. (\ref{Nex}) becomes 
\begin{equation}
\xi=\xi_0+\xi_{\rm ex}=\xi_0 + \ln\left(1+e^{a-\xi_0}\right) \,.
\label{xiaxi0}
\end{equation}
In the canonical ensemble I express $a$ in terms of $\xi$ and $\xi_0$, 
\begin{equation}
a = \xi_0+\ln\left(e^{\xi-\xi_0}-1\right) \,, \label{axixi0}
\end{equation}
and then I write the parition function $\beta\sigma^{-1}\Z_{n_0}$ in terms of 
$\xi$ and $\xi_0$: 
\begin{eqnarray}
\beta\sigma^{-1}\Z_{n_0} &=& (1-h_0)\xi_0^2+\xi_0\ln\left(e^{\xi-\xi_0}-1
\right) + \int_0^\infty\ln\left[1+\left(e^{\xi-\xi_0}-1\right)e^{-x} 
\right] \nonumber \\
&=& (1-h_0)\xi_0^2+\xi_0\ln\left(e^{\xi-\xi_0}-1
\right)- Li_2\left(1-e^{\xi-\xi_0}\right) \label{Zn0xixi0}
\end{eqnarray}

Using Eq. (\ref{Zn0xixi0}) I can find the maximum of 
$\beta\sigma^{-1}\Z_{n_0}$ at fixed $\xi$, and for this 
I equate the derivative 
$\partial(\beta\sigma^{-1}\Z_{n_0})/\partial\xi_0$ to zero, 
\begin{subequations} \label{eqmaxprob}
\begin{equation}
\frac{\partial}{\partial\xi_0}(\beta\sigma^{-1}\Z_{n_0}) = 
(1-2h_0)\xi_0 + \ln\left(1-e^{\xi_0-\xi}\right) - \frac{\xi}{e^{\xi-\xi_0}-1} 
= 0 \,. \label{eqmaxprob1}
\end{equation}
To discuss the solutions of Eq. (\ref{eqmaxprob}), it is better 
to write it in the form 
\begin{equation}
(1-2h_0)\xi_0 = \frac{\xi}{e^{\xi-\xi_0}-1}- \ln\left(1-e^{\xi_0-\xi}\right) 
\label{eqmaxprob2}
\end{equation}
\end{subequations}
and denote the r.h.s. by $f(\xi,\xi_0)$. The first and the second derivative 
of $f(\xi,\xi_0)$, 
%
\begin{eqnarray*}
\frac{\partial f(\xi,\xi_0)}{\partial\xi_0} &=& 
\frac{\xi e^{\xi-\xi_0} + e^{\xi-\xi_0} - 1}{\left(e^{\xi-\xi_0} - 1\right)^2} 
> 0 
\end{eqnarray*}
and 
\begin{eqnarray*}
\frac{\partial^2 f(\xi,\xi_0)}{\partial\xi_0^2} &=& 
\frac{e^{\xi-\xi_0}}{\left(e^{\xi-\xi_0}-1\right)^3}\left(\xi e^{\xi-\xi_0} 
+ e^{\xi-\xi_0} + \xi -1 \right) > 0
\end{eqnarray*}
%
are both bigger than zero for $\xi>0$ and $0\le\xi_0\le\xi$; therefore 
$f(\xi,\xi_0)$, as a function of $\xi_0$, is concave upwards. 

To see if we have any solutions, I evaluate 
$\partial(\beta\sigma^{-1}\Z_{n_0})/\partial\xi_0$ at $\xi_0=0$ and 
$\xi_0=\xi$, and I get 
\begin{subequations} \label{derZn00xi}
\begin{eqnarray}
\left.\frac{\partial(\beta\sigma^{-1}\Z_{n_0})}{\partial\xi_0}\right|_{n_0=0} 
&=& \ln\left(1-e^{-\xi}\right) - \frac{\xi}{e^{\xi}-1} < 0,\ \forall\xi>0 ,
\label{derZn00} 
\end{eqnarray}
and 
\begin{eqnarray}
\left.\frac{\partial(\beta\sigma^{-1}\Z_{n_0})}{\partial\xi_0}\right|_{n_0=N} 
&=& (1-2h_0)\xi + \lim_{\xi_0\to\xi} \left[\ln(\xi-\xi_0) 
- \frac{\xi}{\xi-\xi_0}\right] \nonumber \\ 
&=& -\infty,\ \forall\xi>0 . \label{derZn0xi} 
\end{eqnarray}
\end{subequations}
In other words, the concave up-wards function $f(\xi,\xi_0)$ is 
bigger than the linear function $(1-2h_0)\xi_0$ at both ends of 
the interval $\xi_0\in[0,\xi]$, so, if they are not tangent, they 
must cross each-other either zero times or two times in this interval. 
Therefore the Eqs. (\ref{eqmaxprob}) have either no solutions or have two 
solutions for $\xi_0\in(0,\xi)$. 

We have here the same situation as in the grandcanonical ensemble. 
For small $\xi$, Eq. (\ref{eqmaxprob2}) have no solution. A situation like 
this is depicted in the left-hand plots in Fig. \ref{Zn0_Zn0der} with 
solid lines. As $\xi$ increases, at a certain value $\xi_t$, 
the function $(1-2h_0)\xi_0$ becomes tangent to $f(\xi,\xi_0)$, and 
$\Z_{n_0}$, as a function of $\xi_0$, has an inflection at the tangent 
point, as shown with dashed lines in Fig. \ref{Zn0_Zn0der}. 
For $\xi>\xi_t$, $\Z_{n_0}$ has two maxima, one at $\xi_0=0$ and another 
one at $\xi_0=\xi_{0,2}$ (I use the same notations as in the canonical 
ensemble). As $\xi$ increases above a critical value, 
say $\xi_{cr}$, the maximum formed at $\xi_{0,2}$ becomes dominant (the 
dotted lines in Fig. \ref{Zn0_Zn0der}) and at $\xi=\xi_{cr}$ we have 
a phase transition. 
\begin{figure}
\begin{center}
\resizebox{100mm}{!}{\includegraphics{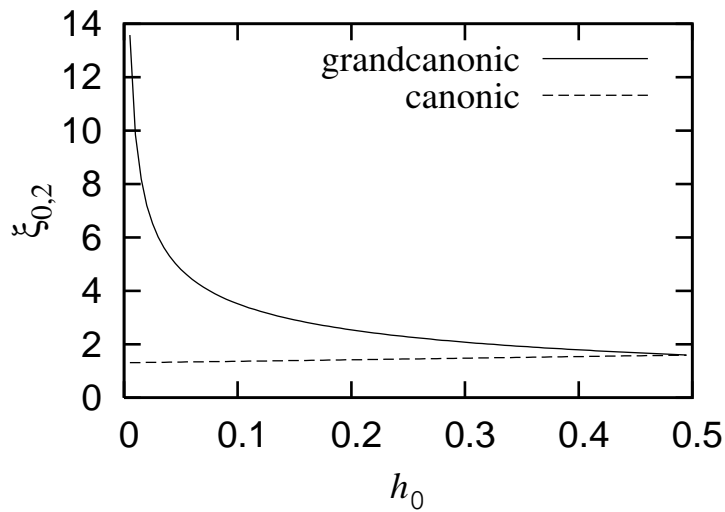}}
\end{center}
\caption{Comparison between of the jumps of the 
number of particles in the condensate at the transition temperature, in 
canonical (dashed line) and grandcanonical (solid line) ensembles.}
\label{comp_g_c_xi02}
\end{figure}

In Fig.~\ref{comp_g_c_xi02} I plot for comparison $\xi_{0,2}$ as a function 
of $h_0$ for the grandcanonical (solid line) and canonical (dashed line) 
ensembles. It is remarcable that although as $h_0\to 0.5$ the Fermi gas 
approaches the ideal behavior, $\xi_{0,2}$ does not converge to zero, 
as one would expect (i.e. no condensation in ideal, continous systems). This 
means that asymptotically, in the noninteracting system a finite fraction 
of the gas condenses at the transition temperature 
\cite{JPA36.L577.2003.Anghel}. 

In Section \ref{DiscC} I shall calculate the latent heat of this transition. 

\section{DISCONTINUITIES AT THE PHASE TRANSITION \label{PhTrDisc}} 

Let us now calculate the jumps of the thermodynamic quantities that 
occure at the phase transition. Usually one expects that a first order 
phase transition in the grandcanonical enemble is marked by 
discontinuities in the internal energy (i.e. a latent heat) and in the 
particle number. In the canonical ensemble, the particle number is held 
fixes and the only quantity that may vary is the internal energy. 
If we think of typical first order phase transitions, like 
liquid-vapour transition or Bose-Einstein condensation, we observe that 
they do not show any discontinuity in the internal energy in canonical 
conditions, but condense gradually as the 
temperature decreases. Unlike these, the transition described here, under 
canonical conditions, is marked by a jump of particles in the condensate, 
accompanied by another jump in the specific heat. 

I shall explain these in the next two subsections. 

\subsection{GRANDCANONICAL ENSEMBLE \label{DiscGC}}

Let us evaluate first the latent heat of the transition in the 
grandcanonical ensemble. If we denote by $U$ the internal energy 
of the system and introduce the adimensional quantity 
$\U(a,\xi_0)\equiv\beta^2\sigma^{-1}U$, by standard calculations I obtain 
\begin{eqnarray}
\U(a,\xi_0) &=& h_0\xi_0^2 + \xi_0\ln 
\left(1+e^{a-\xi_0}\right) + \int_{0}^\infty 
\ln\left(1+e^{a-\xi_0 -x} \right)\,dx \nonumber \\
&=& h_0\xi_0^2+\xi_0\ln\left(1+e^{a-\xi_0}\right) 
- Li_2(-e^{a-\xi_0}) \label{Ugc}
\end{eqnarray}
Substracting (\ref{ln_Zn0_xi0}) form (\ref{Ugc}) and ignoring 
$\beta\sigma^{-1}$, I get 
\begin{eqnarray}
\U(a,\xi_0)-\beta\sigma^{-1}\ln\Z_{n_0} &=& \xi_0\left[
\ln\left(1+e^{\xi_0-a}\right)-(1-2h_0)\xi_0\right] \nonumber \\
&=& -\xi_0\frac{\partial(\beta\sigma^{-1}\ln\Z_{n_0})}{\partial \xi_0} \,,
\label{UminusZ}
\end{eqnarray}
so at the phase transition, $\U(a,\xi_0)=\beta\sigma^{-1}\ln\Z_{n_0}$, 
both at $\xi_0=0$ and $\xi_0=\xi_{0,2}$ (where $\xi_{0,2}$ corresponds 
again to the second maximum of $\ln\Z_{n_0}$). But since 
$\ln\Z_{n_0}(a,\xi_0=0)=\ln\Z_{n_0}(a,\xi_{0,2})$ exactly at the phase 
transition, it implies that $\U(a,0)=\U(a,\xi_{0,2})$ and the grandcanonical 
latent heat, $\lambda_g\equiv (k_BT)^2\sigma[\U(a,\xi_{0,2})-\U(a,0)]$ is 
identically zero. 
\begin{figure}
\begin{center}
\resizebox{100mm}{!}{\includegraphics{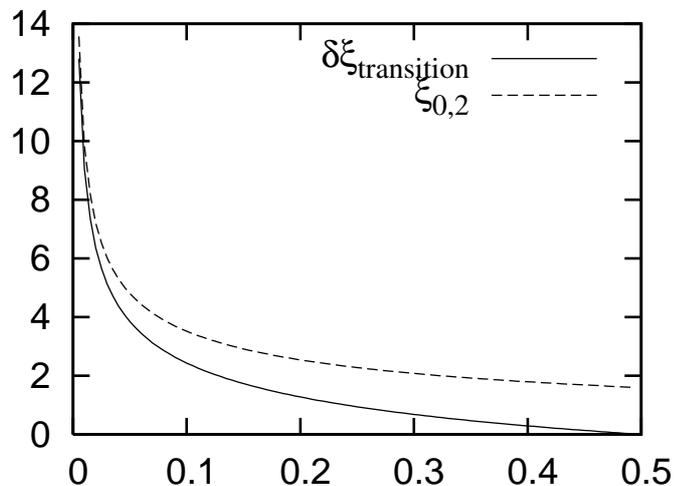}}
\end{center}
\caption{The discontinuity of the particle number in the grandcanonical 
system at the transition temperature (continuous line). With dotted line is 
the }
\label{deltaNg}
\end{figure}

The quantity that varies at the grandcanonical phase transition is the 
particle number. From Eq. (\ref{xiaxi0}) I get directly 
\begin{equation}
\delta\xi_{transition}\equiv\xi(a,\xi_{0,2})-\xi(a,\xi_0=0) = 
\ln\left[\frac{1+e^{\xi_0-a}}{1+e^{-a}}\right]>0 \,. \label{xijump}
\end{equation}
The function $\delta\xi_{transition}$ is plotted in Fig.~\ref{deltaNg}, 
together with the jump of the number of particles in the condensate, 
$\xi_{0,2}$, as a function of $h_0$. Unlike $\xi_0$, $\xi$ converges 
to zero as $h_0$ approaches 0.5. 

\subsection{CANONICAL ENSEMBLE \label{DiscC}}

\begin{figure}
\begin{center}
\resizebox{100mm}{!}{\includegraphics{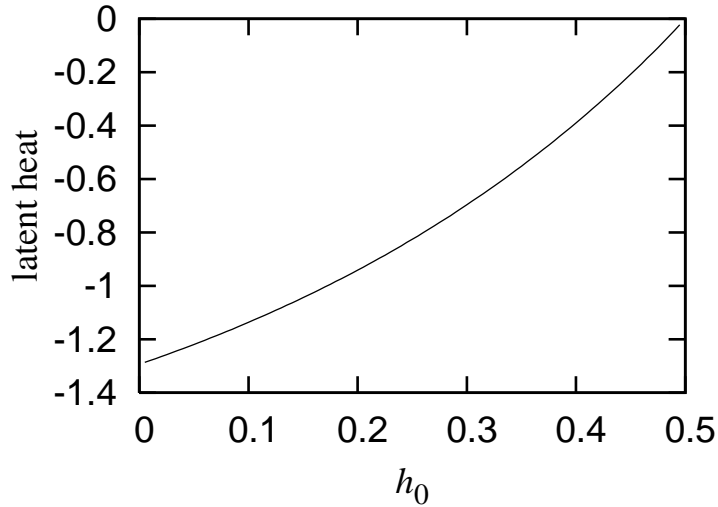}}
\end{center}
\caption{The latent heat of the canonical ensemble, in units 
of $(k_BT)^2\sigma$, as a function of the interaction parameter $h_0$.}
\label{lambdac}
\end{figure}

To calculate the latent heat in the canonical ensemble, I replace 
$a$ in Eq. (\ref{UminusZ}) with the expression (\ref{axixi0}). 
In this way I get 
\begin{eqnarray}
\U(\xi,\xi_0) &=& \xi_0[\xi-\xi_0(1-h_0)] + \int_{0}^\infty 
\ln\left[1+\left(e^{\xi-\xi_0}-1\right)e^{-x}\right]\,dx \nonumber \\
&=& \xi_0[\xi-\xi_0(1-h_0)] - Li_2\left(1-e^{\xi-\xi_0}\right) 
\end{eqnarray}
If I denote by $\lambda_c$ the latent heat in the canonical ensemble, 
then 
\begin{equation}
\frac{\beta^2}{\sigma}\lambda_c = \xi_0[\xi-\xi_0(1-h_0)] + \int_{0}^\infty 
\ln\left[\frac{1+\left(e^{\xi-\xi_0}-1\right)e^{-x}} 
{1+\left(e^{\xi}-1\right)e^{-x}}\right]\,dx \,. 
\end{equation}
This adimensional latent heat is plotted in Fig. \ref{lambdac}.

Therefore the phase transition in the canonical system is marked by 
a discontinuity of the internal energy and a sudden condensation of 
a finite number of particles. 

\section{CONCLUSIONS \label{conclusions}}

I analysed an interacting system model of constant density of single 
particle states, which can describe systems of generalised exclusion 
statistics. I showed that a first order phase transition takes place 
and this phase transition is the condensation of a finite number of 
particles on the lowest energy levels. 
I characterised this transition for the situation when the system can be 
described as a Fermi gas. In grandcanonical conditions, the phase transition 
has no latent heat, but has a discontinuity in the number of particles, 
due to the condensation phenomenon. Unlike usual first order phase 
transitions, in 
this model we have a latent heat at the phase transition in canonical 
ensemble. 

Another interesting fact is that at the transition temperature, the 
number of particles in the condensate converges to a finite, nonzero value, 
as the system is changed gradually into a noninteracting system. This 
can be put into correspondence with the condensation that occures in 
ideal Fermi systems \cite{JPA36.L577.2003.Anghel}.

\end{document}